\documentclass[prb,aps,amssymb,twocolumn,showpacs,floatfix]{revtex4}

\usepackage{graphicx}
\usepackage{dcolumn}
\usepackage{bm}

\begin{document}

\preprint{}

\title{Time- and polarization-resolved optical spectroscopy of colloidal CdSe nanocrystal quantum dots in high magnetic fields}

\author{Madalina~Furis$^{1}$, Jennifer Hollingsworth$^{2}$, Victor I. Klimov$^{2}$, Scott~A.~Crooker$^{1}$}

\affiliation{$^{1}$National High Magnetic Field Laboratory, Los
Alamos, NM 87545}


\affiliation{$^{2}$Chemistry Division, Los Alamos National
Laboratory, Los Alamos, NM 87545}



\date{\today}

\begin{abstract}
In an effort to elucidate the \textit{spin} (rather than
\textit{charge}) degrees of freedom in colloidal semiconductor
nanocrystal quantum dots, we report on a series of static and
time-resolved photoluminescence measurements of colloidal CdSe
quantum dots in ultra-high magnetic fields up to 45 Tesla.  At low
temperatures (1.5 K - 40 K), the steady-state photoluminescence
(PL) develops a high degree of circular polarization with applied
magnetic field, indicating the presence of spin-polarized
excitons.  Time-resolved PL studies reveal a marked decrease in
radiative exciton lifetime with increasing magnetic field and
temperature.  Except for an initial burst of unpolarized PL
immediately following photoexcitation, high-field time-resolved PL
measurements reveal a constant degree of circular polarization
throughout the entire exciton lifetime, even in the presence of
pronounced exciton transfer via F\"{o}rster energy transfer
processes.
\end{abstract}

\maketitle
\textbf{Introduction}

The remarkable optical properties of colloidal semiconductor
nanocrystal quantum dots (NQDs) have attracted considerable
interest across the chemical-, physical-, and materials science
communities in recent years.  Most notably, the zero-dimensional
nature of the quantum-confined electrons and holes in NQDs leads
to discrete, ``atomic-like" energy levels, whose ground-state
emission can be size-tuned over broad ranges of the visible and
infrared spectrum.\cite {alivisatos}  Furthermore, the nearly
perfect and defect-free nanocrystals grown by modern colloidal
chemistry methods virtually eliminate non-radiative exciton
relaxation, such that quantum yields associated with radiative
recombination approach unity. \cite {peng}  These optical
properties, combined with the flexibility to functionalize the
nanocrystal surface, make NQDs suitable for a wide range of
biological and optical applications such as light harvesting
systems \cite {crooker1}, biolabeling \cite {bruchez, clapp},
light emitting diodes \cite {coe, tessler, achermann1, colvin}, or
optical amplification and lasing. \cite {klimov1}
\newline \indent These attributes and applications are based largely on the charge
and energy degrees of freedom of excitons, electrons, and holes in
colloidal NQDs.  In recent years, however, consideration of the
\textit{spin} degrees of freedom of excitons, electrons, and holes
in quantum structures has received significant attention in view
of potential applications in the new scientific fields of
``spintronics" and quantum computation, wherein future generations
of functional devices based on spin orientation and spin
entanglement have been proposed. \cite {divincenzo, imamoglu,
lovett, chen}  Both epitaxially-grown quantum dots as well as
chemically-synthesized colloidal semiconductor quantum dots are
candidate material systems for spin-based devices, due in large
part to the truly zero-dimensional (and therefore discrete) nature
of the quantum-confined exciton states, which leads to long
radiative lifetimes \cite {crooker2} at low temperatures and
extended spin coherence. \cite {gupta}  Further, the enhancement
of the electron-hole exchange interaction in NQDs lifts the
exciton spin degeneracy, such that the exciton ground state in
CdSe NQDs, for example, is characterized by a total spin $J = 2$,
which is optically forbidden, or ``dark", in the dipole
approximation. \cite {efros1, nirmal1, nirmal2, chamarro}  Much
can be learned about these dark excitons through time- and
polarization-resolved optical studies at high magnetic fields,
where dark excitons become mixed with optically allowed
(``bright", $J = 1$) states.  Since significant mixing occurs only
when the exciton Zeeman energy ($E_{Z}=g_{ex}\mu_{B}B$) is
comparable to the bright-dark energy splitting ($\Delta_{bd} >$ 10
meV in small NQDs), magnetic fields on the order of 50 T are
typically required to obtain sufficiently strong mixing between
different spin excitonic states.  Here, $g_{ex}$ is the exciton
Lande g-factor, $\mu_{B}$ is the Bohr magneton, and $B$ is the
magnetic field in Tesla.
\newline \indent In this paper, we present the results of a detailed polarization and
time-resolved magneto-photoluminescence study of colloidal CdSe
NQDs in high magnetic fields to 45 T.  The photoluminescence (PL)
from these nanocrystals is characterized by a high degree of
circular polarization in large magnetic fields, indicating the
presence of spin-polarized excitons.  The measured polarization is
shown to be consistent with excitons that are thermally
distributed amongst Zeeman-split exciton spin levels.  Aside from
an initial burst of unpolarized PL, the measured degree of
circular polarization remains constant throughout the entire
exciton lifetime, even in the presence of strong inter-dot exciton
transfer via the F\"{o}rster energy transfer mechanism.  Exciton
radiative lifetimes are markedly reduced with applied magnetic
field, illustrating the field-induced mixing of the long-lived
dark (spin-2) exciton ground state with higher-energy,
optically-allowed (spin-1) exciton states.

\textbf{Experimental Section}

\textbf{Nanocrystal Synthesis.} The data presented in this paper
are derived from three NQD samples of different size.  Colloidal
CdSe NQDs of 26 \AA~  and 40 \AA~  diameter were grown by
organo-metallic synthesis, following the methods of Murray and
co-workers. \cite {murray}  The cation and anion precursors,
Me$_{2}$Cd and (TMS)$_{2}$Se, were combined and injected in a
heated (200 $^{o}C$) trioctylphosphine oxide (TOPO) solution. The
reaction mixture is heated to temperatures varying between 230
$^{o}C$ and 260 $^{o}C$ for a few hours.  These two NQD samples
were overcoated with ZnS through a similar reaction for improved
surface passivation and increased PL quantum yield.  In addition
to acting as a solvent, the TOPO also served as a surfactant,
providing solubility.  Several precipitations of the NQDs with
methanol were carried out in order to remove all excess unbound
ligand, and the NQDs were then re-dispersed in hexane.  A third
sample of as-grown 57\AA~  diameter NQDs (no ZnS overcoat) was
also prepared through similar methods.  For the experiments in
high magnetic fields, the samples were prepared by drop-casting
NQDs from hexane/octane solutions onto glass slides, forming high
quality, optically clear NQD films.
\newline \indent \textbf{Static Photoluminescence in Pulsed Magnetic Fields.} PL spectra in high
magnetic fields were acquired at the National High Magnetic Field
Laboratory (Los Alamos), using a 50 T pulsed magnet.  This magnet
provides pulses of roughly 300 millisecond duration, and is
powered by a 1.6 megajoule capacitor bank (10 kVolt, 32
milliFarads).  For studies at temperatures between 1.5 K and 4 K,
the NQD samples were mounted on a fiber-optic probe and loaded
directly in the (pumped) liquid volume of a helium bath cryostat.
 For temperatures above 4 K, the sample probe was loaded into an
additional vacuum jacket and backfilled with helium exchange gas.
 A single 600 micron diameter optical fiber couples 442 nm
excitation light from a helium-cadmium laser (Kimmon Electric Co.)
to the sample.  The same fiber also serves to collect the emitted
PL and direct it to a spectrometer.  Thin-film circular polarizers
sandwiched between the fiber and the sample permit polarization
analysis of the PL.  The collected PL is dispersed in a 300 mm
spectrometer (Acton 308) and detected with a high resolution,
liquid-nitrogen cooled CCD camera capable of continuous
acquisition of spectra at rates up to 1 kHz (Princeton
Instruments, 1340x100 pixel array).  In this way, the complete
magnetic field dependence of the PL is acquired during each magnet
pulse, with roughly 1 hour between magnet pulses. \cite {crooker3}
\newline \indent \textbf{Time-Resolved Photoluminescence in dc
Magnetic Fields.}  Time-resolved PL measurements were performed in
the variable-temperature insert of an 18 T superconducting magnet
(Oxford Instruments), using techniques and hardware for
time-correlated single-photon counting (Picoquant SPC-430). Here,
the NQDs were excited at 410 nm by a frequency-doubled ultrafast
Ti:sapphire laser (Coherent Mira 900) equipped with an external
pulse-picker to reduce the repetition rate. \cite {crooker1}  The
PL was spectrally filtered in a 275 mm spectrometer (Acton 275) to
a narrow ($< 1$ nm) bandwidth, and was detected with a
multichannel-plate photomultiplier tube (Hamamatsu R3809U-51).
 Modal dispersion in the long (25 meter) multimode optical fiber
limited the system time resolution to $\sim 500$ ps.

\textbf{Results and Discussion}

\textbf{Exciton ``Fine Structure" in NQDs.}  Experimental and
theoretical studies have established that the enhanced
electron-hole exchange interaction in wurtzite CdSe NQDs lifts the
spin degeneracy of the band-edge exciton. \cite {efros1, nirmal1,
nirmal2, chamarro}  The result, shown in Figure 1a, is a
five-level exciton ``fine structure", in which the five states are
characterized by their total spin projection $J$ along the
symmetry-breaking crystalline \textbf{c}-axis of the wurtzite NQD.
\cite {efros1, efros2} In prolate or spherical NQDs, the lowest
(ground state) exciton has spin projection $J = 2$, which
therefore cannot couple directly to light (photons having spin
equal to 1).  This $J = 2$ dark exciton resides 2-18 meV
(depending on NQD size) below the lowest optically allowed ($J =
1^{L}$), bright exciton state. \cite {nirmal2}  The evolution of
this bright-dark energy gap, $\Delta_{bd}$, as a function of NQD
size can be seen in Figure 1a.
 Within this framework and in the absence of external magnetic
fields, previous studies \cite {crooker2} have shown that the
radiative recombination of $J = 2$ dark excitons occurs either via
thermal activation to the lowest optically-allowed state ($J =
1^{L}$), or at the lowest temperatures, through higher-order
processes such as LO phonon assisted transitions. (In this case,
the LO phonon carries away the extra unit of angular momentum).
\begin{figure}[tbp]
\includegraphics[width=.43\textwidth]{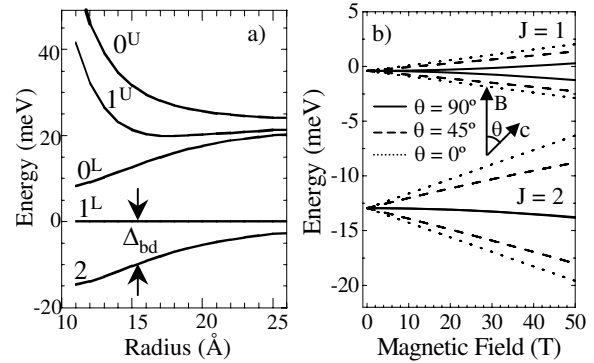}
\caption{ (a) ``Fine structure" of the ground state exciton in
CdSe nanocrystal quantum dots (NQDs) as a function of dot radius.
The energy levels are labelled by the net spin projection J along
the symmetry-breaking crystalline \textbf{c}-axis of the wurtzite
NQD. (from ref. 17) (b) Calculated Zeeman splitting of the two
lowest exciton levels ($J = \pm 2$, and $J = (\pm 1)^{L}$) in CdSe
NQDs as a function of magnetic field $B$, for three orientations
of the nanocrystal \textbf{c}-axis with respect to the applied
field ($\theta$ = 0, 45, and 90 degrees). (following ref. 23)}
\label{fig1}
\end{figure}
\newline \indent Application of an external magnetic field allows for direct
radiative recombination of dark excitons, due to field-induced
mixing between the dark and bright exciton states. \cite {nirmal2}
 As described by Efros \cite {efros2}, the degree of bright-dark
mixing depends critically on the orientation of the applied
magnetic field with respect to the crystalline \textbf{c}-axis of
the NQD.  In zero magnetic field, exciton spins are naturally
aligned (and quantized along) the \textbf{c}-axis of the wurtzite
NQD.  Magnetic fields exactly parallel to the \textbf{c}-axis
cause no bright-dark mixing, and the \textbf{c}-axis remains a
good spin quantization axis.  Parallel magnetic fields result in
the usual Zeeman spin splitting of  $J = (\pm 1)^{L}$  and $J =
\pm 2$ excitons.  In contrast, magnetic fields orthogonal to the
NQD \textbf{c}-axis strongly mix the lowest $J = \pm 2$ dark
states with the optically active $J = (\pm 1)^{L}$ excitons
(\textit{i.e.}, the \textbf{c}-axis is no longer a valid spin
quantization axis).  This mixing allows PL emission directly from
the lowest exciton states, leading to a faster exciton decay even
at low temperature.  No Zeeman splitting of the $J = \pm 2$
exciton states is induced, to first order, for fields exactly
orthogonal to the \textbf{c}-axis.  In general (see Figure 1b),
for an NQD having its \textbf{c}-axis at an angle $\theta$ with
respect to the applied magnetic field, the parallel component of
the magnetic field, $B \cos \theta$, generates a Zeeman splitting
of the states, while the perpendicular component of the magnetic
field, $B \sin \theta$, induces mixing between the dark and bright
states.  In an actual ensemble, the NQDs have random orientation,
so that quantitative interpretation of magneto-optical data
requires proper averaging over all possible NQD orientations.
\newline \indent \textbf{Circularly-Polarized Photoluminescence in
Ultrahigh Magnetic Fields}.  Figure 2 shows characteristic
polarization-resolved PL spectra from NQDs at low temperatures.
 These data show the 1.6 K PL spectra from the sample of 26\AA~
NQDs, in both right ($\sigma^{+}$) and left ($\sigma^{-}$)
circular polarizations, at 0, 15, 30, and 45 T collected in the
Faraday geometry (\textbf{B} parallel to the incident light
direction).  At this temperature, the zero-field PL from these
small NQDs is peaked at 2.41 eV and has a full width at
half-maximum (FWHM) of $\sim 100$ meV, due to the $5\%$ dispersion
of NQD size in the sample.  With increasing magnetic field, the
data show a significant degree of circular polarization.  The
intensity of the $\sigma^{-}$ emission increases steadily with
applied magnetic field, while emission from the opposite circular
polarization ($\sigma^{+}$) decreases with field and remains
roughly constant above 15 T.  The marked circular polarization
that develops with applied magnetic field indicates the presence
of predominantly spin-polarized excitons in the NQDs.  Only small
changes in the PL linewidth are observed.  Further, although
magnetic fields dramatically reduce the observed PL lifetimes (to
be shown later), the time- and polarization-integrated PL emission
varies only by $\sim 10\%$ as a function of magnetic field - again
indicating the nearly complete absence of competing nonradiative
recombination channels in these NQDs.
\begin{figure}[tbp]
\includegraphics[width=.43\textwidth]{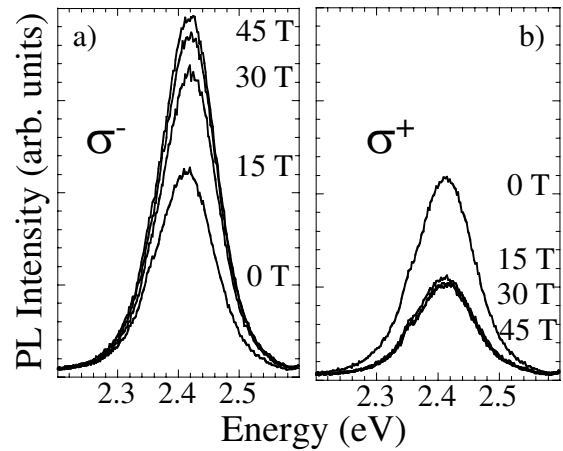}
\caption{ Photoluminescence (PL) spectra acquired at 1.6 K from
the 26 \AA~  diameter CdSe/ZnS NQDs at 0, 15, 30, and 45 T in both
(a) $\sigma^{-}$ and (b) $\sigma^{+}$ circular polarizations.}
\label{fig2}
\end{figure}
\newline \indent The complete field- and polarization-dependent intensities for the
same 26\AA~  NQDs are shown in Figures 3a-d, at 1.6, 4.0, 10, and
20 K.  At the lowest temperatures (Figure 3a), the intensity of
the $\sigma^{-}$ emission grows steadily, increasing by a factor
of 1.75 at the maximum applied magnetic field of 45 T, while the
intensity of the $\sigma^{+}$ emission decreases to half its
initial value by 10 T and remains roughly constant thereafter. At
higher temperatures to 20 K, these intensity shifts remain
qualitatively the same, but are reduced somewhat in magnitude. The
corresponding peak energies of the PL are shown in Figures 3e-h.
At higher temperatures (4 K - 20 K), the zero-field PL is peaked
at 2.419 eV.  With applied field, both $\sigma^{+}$ and
$\sigma^{-}$ emission peaks exhibit a monotonic redshift toward
lower energies, with the $\sigma^{+}$ emission showing a larger
energy shift, leading to a small but apparent splitting between
the two opposite circular polarizations.  At the lowest
temperature (\textit{i.e.}, the 1.6 K data of Figure 3e), the
zero-field PL appears at slightly lower energy (2.411 eV), due to
an increased probability of phonon-assisted emission from these
dark excitons. \cite {nirmal2}  The subsequent non-monotonic
energy shifts observed between 0 and 20 T result from the
field-induced mixing of dark excitons with higher-energy
bright-exciton states, increasing  the likelihood of direct
recombination without involvement of phonons.
\begin{figure}
\includegraphics[width=.45\textwidth]{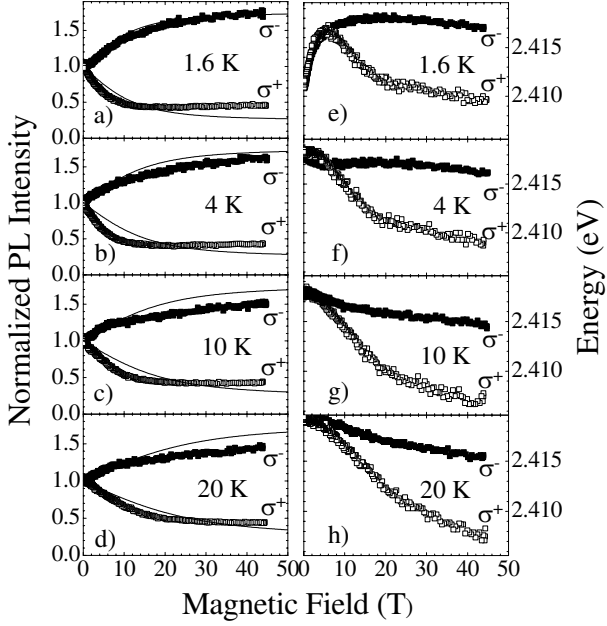}
\caption{(a-d) Integrated intensities of the $\sigma^{-}$ and
$\sigma^{+}$ PL emission as a function of applied magnetic field,
at the temperatures indicated, in the 26 \AA~  CdSe/ZnS NQDs. The
continuous lines represent the expected evolution of the
$\sigma^{-}$ and $\sigma^{+}$ components of the emission from
randomly-oriented NQDs containing a thermal population of $J = -2$
and $J = +2$ dark excitons as described by Equations (1) and (2).
(e-h) Corresponding energy of the PL emission peak, showing
splitting between $\sigma^{-}$ and $\sigma^{+}$ components, and
markedly non-monotonic behavior at the lowest temperatures.}
\label{fig3}
\end{figure}
\newline \indent It is notable that the intensity of the
$\sigma^{+}$ PL emission does not approach zero at high magnetic
fields, but rather saturates at a finite value (see Figures 3a-d).
 This behavior is a direct result of the random orientation of NQD
\textbf{c}-axis (with respect to the applied magnetic field) in
the sample.  Following the ideas of Efros and Johnston-Halperin
\cite {efros2, johnstonhalperin}, this data can be qualitatively
modelled by considering circularly-polarized emission from
randomly-oriented NQDs containing a thermal population of $J =\pm
2$ dark excitons.  In the present Faraday geometry, spin-polarized
excitons in an NQD with the \textbf{c}-axis oriented at an angle
$\theta$ to the applied field will emit $\sigma^{+}$ light with
intensity proportional to $ (1+\cos\theta)^{2}$, and $\sigma^{-}$
light with intensity proportional to $ (1-\cos\theta)^{2}$.  For
oppositely-oriented excitons, these intensities are reversed:
$(1-\cos\theta)^{2}$ for $\sigma^{+}$ light, and $
(1+\cos\theta)^{2}$ for $\sigma^{-}$ light.  Note, therefore, that
NQDs with \textbf{c}-axis parallel to the applied field ($\theta =
0$) emit $100\%$ circularly polarized light along the direction of
observation, but that NQDs with \textbf{c}-axis orthogonal to the
applied field ($\theta  = 90$ degrees) emit unpolarized light
along the direction of observation.  Boltzmann (thermal)
statistics determines the relative population of $J = \pm 2$
excitons, given the orientation-dependent $J = \pm 2$ Zeeman spin
splitting $E_{Z}=g_{ex}\mu_{B}B\cos\theta$, where $g_{ex}$ is the
dark exciton g-factor.  Letting $x=\cos\theta$ and
$\beta=(k_{B}T)^{-1}$, the angle-dependent intensities of emitted
$\sigma^{-}$ and $\sigma^{+}$ light are:
\begin{equation}
I_{\sigma^{-}}(x)=1+x^{2}+2x\tanh(E_{Z}\beta/2)
\end{equation}
\begin{equation}
I_{\sigma^{+}}(x)=1+x^{2}-2x\tanh(E_{Z}\beta/2)
\end{equation}
Integrating $I_{\sigma^{-}}$ and $I_{\sigma^{+}}$ over all NQD
orientations ($0 < x < 1$) gives the total PL intensities, which
are shown for comparison by the lines in Figure 3a.  Note that the
predicted $\sigma^{+}$ intensity saturates at a finite value even
in the limit of high magnetic fields.  Imperfect agreement between
the measured data and the straightforward model outlined above
likely originates in the exact role played by phonon-replica
emission (\textit{i.e.} the contribution of the phonon-replicas to
the PL spectrum), and the strong dependence of the dark exciton
recombination rate on the magnetic field as a result of the mixing
with  $J = 1$ bright states.  While the overall trends and
significant features in the field-dependent intensity may be
understood within this model, additional theoretical efforts are
required to elucidate the differences.
\newline \indent From the data of Figures 3a-d, the degree of circular polarization of the PL emission, $P=(I_{\sigma^{-}}-I_{\sigma^{+}})/(I_{\sigma^{-}} +
I_{\sigma^{+}})$, can be computed as a function of magnetic field.
 This polarization is shown in Figure 4 at 1.6 K, 4.0 K, 10 K, and
20 K for all three sizes of NQDs.  The finite intensity of
$\sigma^{+}$ emission (even at the highest magnetic fields) leads
to saturation of the degree of circular polarization at a value
less than $100\%$.  The temperature dependence of the polarization
also follows the behavior of a Boltzmann population of dark
excitons, similar to previous studies. \cite {johnstonhalperin}
\begin{figure}
\includegraphics[width=0.33\textwidth]{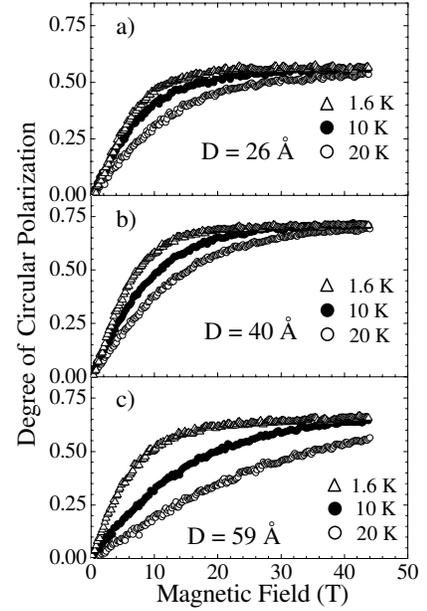}
\caption{Degree of circular polarization versus applied magnetic
field at 1.6 K, 10 K, and 20 K in the (a) 26 \AA~  CdSe/ZnS NQDs,
(b) 40 \AA~  CdSe/ZnS NQDs, and (c) the 57 \AA~  CdSe/TOPO NQDs.}
\label{fig4}
\end{figure}
\newline \indent \textbf{Time-Resolved Photoluminescence Studies in High
Magnetic Fields.}  We turn now to explicit measurements of the
magnetic field dependence of recombination dynamics in CdSe NQDs.
 As previously established \cite {crooker2, nirmal1, nirmal2}, the
measured radiative lifetime ($\tau_{R}$) of excitons in NQDs
becomes very long - of order of 1 $\mu$s -- at low temperatures
below 4 K, as compared with $\tau_{R}\sim$ 20 ns at room
temperature.  This dramatic increase in $\tau_{R}$ is due to the
fact that, below 4 K, excitons are largely frozen out in the $J =
2$ dark exciton ground state, from which direct radiative
recombination is forbidden to leading order.  As discussed above,
these dark excitons eventually undergo radiative recombination
through phonon-assisted (or other weakly allowed) processes, in
which the phonon takes away the additional quantum of angular
momentum.  These slow ``higher-order" recombination processes
ultimately limit $\tau_{R}$ to 1 $\mu$s at low temperatures.
\newline \indent While $J = 2$ exciton states also exist in bulk semiconductors
and 2-dimensional quantum well systems, the energy splitting
between dark and bright states is typically very small
($\Delta_{bd} \sim 0.1$ meV) \cite {blackwood}, so that even at
very low temperatures of order 1 K, the thermal energy, $k_{B}T
\sim 0.09$ meV, is comparable to $\Delta_{bd}$.  Thus, thermal
excitation of dark excitons to short-lived bright exciton states
dominates, and the exciton lifetime in bulk or 2D semiconductors
remains fast.  In contrast, the bright-dark splitting in NQDs is
quite large ($\Delta_{bd} \sim 2-18$ meV), so that appreciable
thermal excitation of dark to bright excitons, and a
correspondingly fast radiative lifetime, is observed \cite
{crooker2} only when the thermal energy $k_{B}T$ approaches
$\Delta_{bd}$ (practically, at temperatures of a few Kelvin).
\begin{figure}
\includegraphics[width=.33\textwidth]{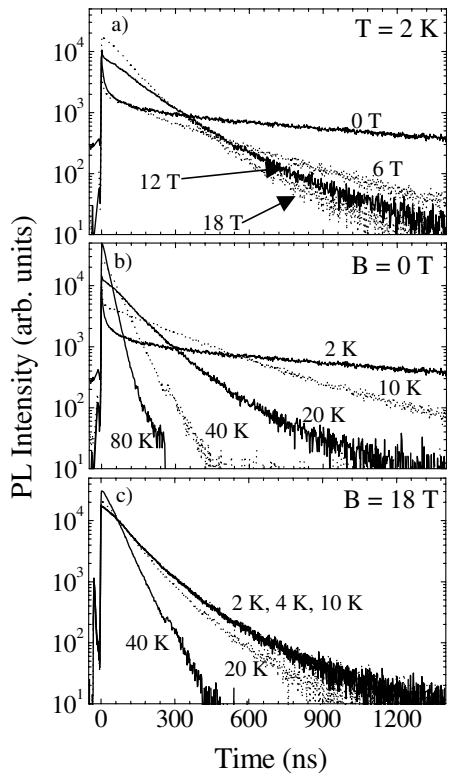}
\caption{Time-resolved PL decays (log scale) from the 26\AA~
CdSe/ZnS NQDs, for (a) different applied fields at 2 K, (b)
different temperatures at 0 T, and (c) for different temperatures
at 18 T (here, note that the lifetime does not change until 20
K).} \label{fig5}
\end{figure}
\newline \indent In analogy with the case of increased temperature, applied
magnetic fields also accelerate the radiative recombination of
dark excitons due to field-induced mixing of dark and bright
exciton states for those NQDs whose \textbf{c}-axes are not
exactly aligned along \textbf{B}. \cite {nirmal2, efros2,
johnstonhalperin}  This effect becomes appreciable when the
magnetic (Zeeman) energy $g_{ex}\mu_{B}B$ approaches $\Delta_{bd}$
(practically, at fields of a few Tesla).  Thus, the effect of
applied magnetic field on the PL dynamics of NQDs is expected to
be qualitatively similar to the effect of increased temperature.
This comparison is shown explicitly in Figure 5. Figure 5a shows
PL decays from the 26 \AA~  dots at low temperature (2 K), at 0,
6, 12, and 18 Tesla.  With increasing magnetic field, the PL
lifetime becomes shorter, decreasing from $\sim$ 1 $\mu$s at 0 T
to $\sim 100$ ns at 18 T, consistent with field-induced mixing of
dark and bright exciton states.  We note that the measured PL
decays in applied magnetic field are expected to be
non-exponential, as they contain contributions from
randomly-oriented NQDs (each particular NQD orientation gives a
particular amount of bright-dark mixing, and thus a particular
radiative lifetime).  Nonetheless, the data of Figure 5a show that
the PL dynamics at long delays (hundreds of nanoseconds) are
largely exponential, and for the purposes of empirical comparison,
can be characterized by a single radiative lifetime.
\begin{figure}[tbp]
\includegraphics[width=.43\textwidth]{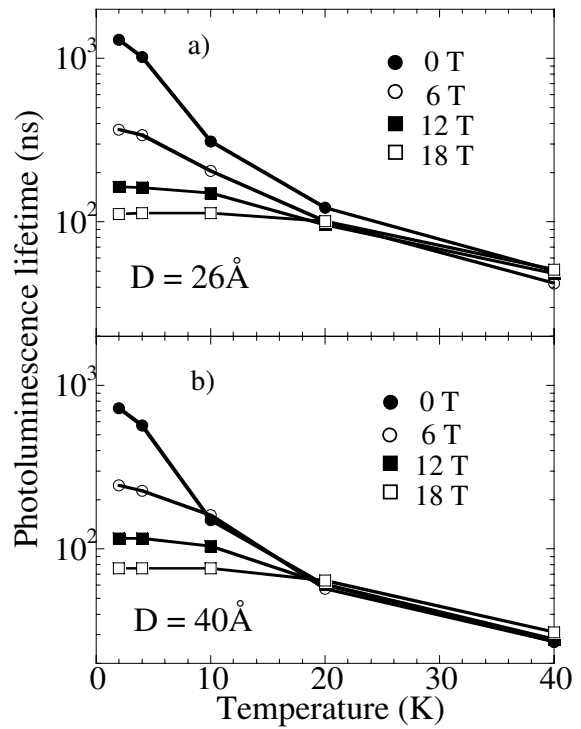}
\caption{ (a) Measured PL lifetimes versus temperature, at
different magnetic fields in (a) 26 \AA~ CdSe/ZnS NQDs and (b) 40
\AA~ CdSe/ZnS NQDs.} \label{fig6}
\end{figure}
  Figure 5b shows the similar effect of increased temperature on PL
dynamics. Here, at zero applied magnetic field, increasing
temperature accelerates the PL radiative decay due to thermal
activation of dark to bright exciton states, as discussed above
and observed previously. \cite {crooker2}  These data show that
the lifetime is reduced from $\sim$ 1 $\mu$s at 2 K to $\sim 100$
ns at 20 K, similar to the effect of an 18 T applied magnetic
field. When \textit{both} magnetic fields and elevated
temperatures are present, the PL radiative lifetime is therefore
limited primarily by the \textit{larger} of the two energy scales
($k_{B}T$  or $g_{ex}\mu_{B}B$), as shown in Figure 5c.  These
data show that at 18 Tesla, the PL lifetime remains unchanged by
increases in temperature, until it rises above a temperature of
$\sim 20$ K. The interplay between the effects of temperature and
applied field on PL lifetime are summarized in Figure 6, where the
measured PL lifetimes are shown for the 26 \AA~  and 40 \AA~ NQDs
as a function of both temperature and magnetic field.  As can be
seen, application of an 18 T magnetic field at the lowest
temperatures reduces the radiative lifetime by an order of
magnitude, whereas increasing the temperature while in the
presence of high magnetic fields has little effect on the PL
lifetime.  We note again that the time- and
polarization-integrated PL intensity remains virtually unchanged
by increasing magnetic field or increasing temperature (despite
the large changes in PL lifetime), indicating the near-absence of
competing non-radiative recombination processes.
\newline \indent \textbf{Dynamic Circular Polarization.}  Immediately following
excitation of NQDs by the pulsed laser, the photo-injected
excitons undergo rapid (sub-picosecond to picosecond timescale)
energy relaxation down to the band-edge \cite {klimov2}, and
distribute themselves among the Zeeman-split spin states of the
lowest exciton level.  The static, circularly-polarized PL
measurements shown in the previous section are consistent with a
Boltzmann thermal distribution of excitons.  Time-resolved
polarized PL measurements, shown in Figure 7, demonstrate that the
degree of PL circular polarization (and therefore exciton spin
polarization) is constant throughout the long exciton lifetime.
 Figure 7 shows decays of the $\sigma^{+}$ and $\sigma^{-}$ PL
components from the 26 \AA~ NQDs at B = 18 T, at both 2 K and 20
K. PL decays at 0 T are also shown for comparison.  This
observation suggests that the process of exciton thermalization
between states with different spins occurs on a timescale that is
considerably shorter than the exciton lifetime.
\newline \indent At very early time delays ($< 1$ ns), immediately following
photoexcitation, the data do reveal a fast, unpolarized component
of the PL emission.  Figure 8a shows polarization-resolved PL
decays at 2 K and at 12 T, where the $\sigma^{+}$ and $\sigma^{-}$
emission intensities are initially nearly equal, before settling
down to a constant degree of polarization.  This initial burst of
unpolarized emission disappears with increasing temperature, as
shown in Figure 8b, where the relative amplitude of the initial
burst is diminished by 20 K and absent at 40 K.
\begin{figure}[tbp]
\includegraphics[width=.43\textwidth]{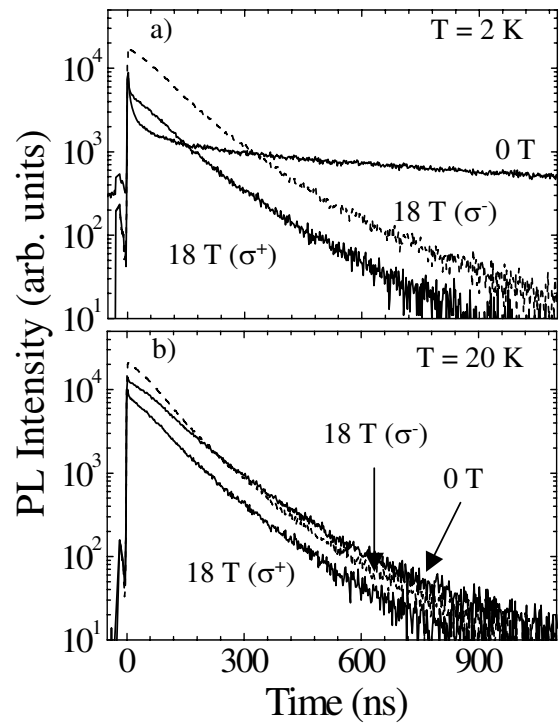}
\caption{ (Time- and polarization-resolved PL decays in the 26
\AA~ CdSe/ZnS NQDs at 18 T, at (a) 2 K and (b) 20 K, showing
constant degree of circular polarization throughout the exciton
lifetime} \label{fig7}
\end{figure}
  The sub-nanosecond measured timescale of this initial decay is
commensurate with the 500 ps system time resolution; thus, it is
likely that the dynamics of this unpolarized component are faster.
This burst of emission may result from the relaxation of the
non-thermal and unpolarized excitons that are photoinjected at
$\sim 3.0$ eV by the pulsed, frequency-doubled Ti:Sapphire laser.
 As these unpolarized excitons decay in energy towards the lowest
dark exciton levels at 2.4 eV, they must relax through the cascade
of higher-lying optically-allowed levels, from which the
probability for radiative recombination is much larger.  Further
experiments with improved time resolution are necessary to resolve
the timescale of exciton spin thermalization within the manifold
of band-edge exciton states.
\newline \indent \textbf{Energy Transfer in Magnetic Fields.}  Finally, recent experiments have established that inter-dot communication via
F\"{o}rster energy transfer is an effective means of NQD coupling.
\cite {crooker1, achermann2, kagan}  These studies portend
favorably for an efficient means of energy and exciton transfer,
in applications such as light harvesting, using artificial
materials constructed of colloidal NQDs or ``non-contact"
activation of NQDs using exciton transfer from proximal epitaxial
semiconductor layers. \cite {achermann1}
\begin{figure}[tbp]
\includegraphics[width=.45\textwidth]{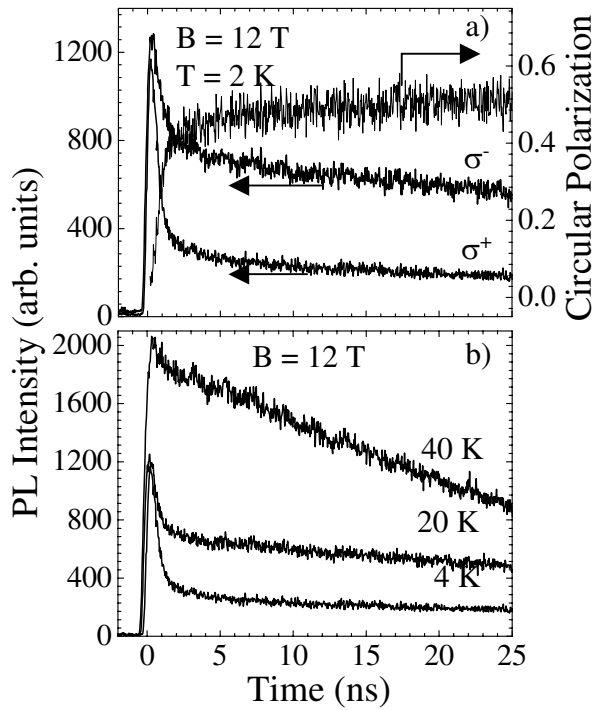}
\caption{ (a) Expanded view of the polarization-resolved initial
PL decay and the circular polarization decay in 26 \AA~  dots at
12 T, showing an unpolarized initial spike at low temperatures (2
K). The width of initial spike is limited by the system response
time. (b) Temperature evolution of the initial PL decay, showing
that the initial burst of unpolarized PL disappears with increased
temperatures.} \label{fig8}
\end{figure}
  We show in this last section that inter-dot coupling via F\"{o}rster transfer processes
may also preserve exciton spin degrees of freedom; an important
characteristic for future ``spintronic" devices and applications
of quantum computing.  Following the measurements of ref. 3,
Figure 9a shows time-resolved PL decays at selected wavelengths
spanning the inhomogeneously broadened PL emission line, at an
intermediate temperature of 40 K.  Here, the NQDs in the drop-cast
film are close-packed, so that they are sufficiently close
together to allow for efficient F\"{o}rster transfer between dots.
When measuring the PL decay on the high energy (short wavelength)
side of the PL emission at 2.51 eV, the lifetime is observed to be
fast - of order of 20 ns. Emission at this high energy comes from
the smallest dots in the ensemble, which have largest band gap. As
the detection energy is decreased (moving to longer wavelengths),
the measured PL decay becomes longer and longer, eventually
reaching a value of $\sim 60$ ns at 2.31 eV.
\begin{figure}[tbp]
\includegraphics[width=.43\textwidth]{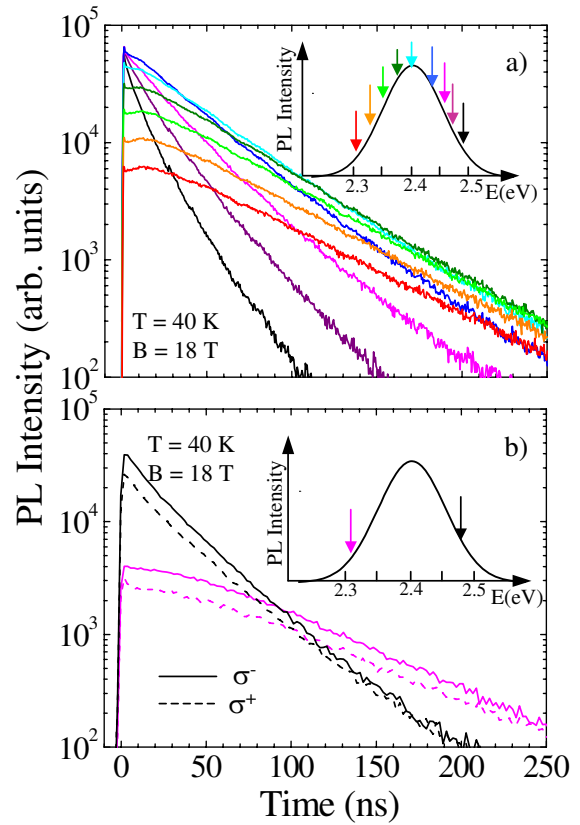}
\caption{ (a) Time- and spectrally-resolved unpolarized PL decays
at selected energies within the PL emission line at 40 K, showing
evidence of F\"{o}rster energy transfer from small NQDs to large
NQDs (fast decays at high energies, slow decays at low energies).
(b) Polarization-resolved PL decays at 2.47 eV and 2.31 eV,
showing that the degree of exciton spin polarization remains
constant even in presence of strong F\"{o}rster energy transfer.}
\label{fig9}
\end{figure}
PL at these longer wavelengths (from the larger dots in the
ensemble) exhibit decidedly non-exponential decays, even showing a
slight increase at short time delays.  As discussed previously
\cite {crooker1} , these data result from F\"{o}rster transfer of
excitons from the smaller (donor) NQDs to the larger (acceptor)
NQDs in the ensemble.  Both the rapid PL decay at high energy, and
the slow non-exponential decay at low energy indicates a flow of
excitons from small to large NQDs occurring after photoexcitation.
In high magnetic fields where the PL emission is circularly
polarized, polarization analysis of the PL dynamics in the
presence of strong F\"{o}rster coupling reveals that the observed
degree of circular polarization is unaffected by the exciton
transfer process.  Figure 9b shows that, at 18 T, the degree of
circular polarization from the larger ``acceptor" NQDs remains
essentially unchanged throughout the exciton lifetime, even though
the overall shape of the decay indicates that a significant number
of excitons are being transferred to these acceptor dots (from
donor dots) well after the initial photoexcitation at $t = 0$.
These data are consistent with a spin-preserving F\"{o}rster
energy transfer process (in which, \textit{e.g.}, a spin-down
exciton in the donor NQD is transferred to a spin-down state in
the acceptor NQD).  However, because the exciton lifetime is much
greater than the spin thermalization time, it is important to note
that these data do not rule out the possibility that F\"{o}rster
energy transfer is spin-incoherent (and excitons in the acceptor
dots become spin polarized by the usual thermalization between
Zeeman split states).  Nonetheless, recent theoretical work by
Govorov \cite {govorov} suggests that spin-coherent transfer of
excitons between quantum dots is expected in randomly-oriented
quantum dots, particularly if resonance conditions between the
donor and acceptor dots are satisfied.  As recently proposed,
spin-preserving exciton-transfer mechanisms can be exploited for
generating quantum entangled states in nanocrystals \cite {lovett,
ouyang}, as long as the energy transfer is fast enough so that
switching between the entangled states occurs on a much faster
timescale than the exciton decoherence times.

\textbf{Conclusions}

In conclusion, the spin polarization of excitons in colloidal CdSe
NQDs is studied via polarization analysis of the static and
time-resolved PL.  High magnetic fields lead to a significant
degree of circularly polarized PL emission up to $70\%$, which can
be modeled by averaging emission from a thermal distribution
between $J = +2$ and $J = -2$  dark excitons in an ensemble of
randomly oriented wurtzite NQDs.  Time-resolved studies show that
magnetic field induced mixing of dark and bright exciton states
leads to markedly reduced exciton lifetimes, quite similar to the
effects of thermal excitation from dark to bright exciton states
due to increased temperature.  Thermalization of photoinjected
excitons between the Zeeman-split $J = \pm 2$ dark exciton states
occurs on sub-nanosecond timescales, after which the degree of
spin polarization remains constant throughout the exciton
lifetime.  This constant degree of spin polarization is robust
even in the presence of strong inter-dot coupling due to
F\"{o}rster exciton transfer processes.


\begin{references}

\bibitem{alivisatos} Alivisatos, A. P. \textit{Science} \textbf{1996}, \textit{271}, 933.

\bibitem{peng} Peng, X.; Schlamp, M. C.; Kadavanich, A. V.; Alivisatos, A. P. \textit{J. Am. Chem. Soc.} \textbf{1997}, \textit{119} 7019.

\bibitem{crooker1} Crooker, S. A.; Hollingsworth, J. A.; Tretiak, S.; Klimov, V. I. \textit{Phys. Rev. Lett.} \textbf{2002}, \textit{89} 186802.

\bibitem{bruchez} Bruchez, M.; Moronne, M.; Gin, P.; Weiss, S.; Alivisatos, A. P. \textit{Science} \textbf{1998}, \textit{281} 2013.

\bibitem{clapp} Clapp, A. R.; Medintz, I. L.; Mauro, J. M.; Fisher, B. R.; Bawendi, M. G.; Mattoussi, H. \textit{J. Am. Chem. Soc.} \textbf{2004}, \textit{126} 301.

\bibitem{coe} Coe, S.; Woo, W. -K.; Bawendi, M.; Bulcovi\'{c}, V. \textit{Nature} \textbf{2002}, \textit{420} 800.

\bibitem{tessler} Tessler, N.; Medvedev, V.; Kazes, M.; Kan, S.; Banin, U. \textit{Science} \textbf{2002}, \textit{295} 1506.

\bibitem{achermann1} Achermann, M.; Petruska, M.; Kos, S.; Smith, D. L.; Koleske, D. D.; Klimov, V. I. \textit{Nature} \textbf{2004}, \textit{429} 642.

\bibitem{colvin} Colvin, V. L.; Schlamp, M. C.; Alivisatos, A. P. \textit{Nature} \textbf{1994}, \textit{370} 354.

\bibitem{klimov1} Klimov, V.I.; Mikhailovsky, A. A.; Xu, S.; Malko, A.; Hollingsworth, J. A.; Leatherdale, C. A.; Eisler, H. -J.; Bawendi, M. G. \textit{Science} \textbf{2000}, \textit{290} 314.

\bibitem{divincenzo} DiVincenzo, D. P.; Bacon, D.; Kempe, J.; Burkard, G.; Whaley, K. B. \textit{Nature} \textbf{2000}, \textit{408} 339.

\bibitem{imamoglu} Imamoglu, A.; Awschalom, D. D.; Burkard, G.; DiVincenzo, D. P.; Loss, D.; Sherwin, M.; Small, A. \textit{Phys. Rev. Lett.} \textbf{1999}, \textit{83} 4204.

\bibitem{lovett} Lovett, B. W.; Reina, J. H.; Nazir, A.; Briggs, G. A. D. \textit{Phys. Rev. B} \textbf{2003}, \textit{68} 205319.

\bibitem{chen} Chen, G.; Bonadeo, N. H.; Steel, D. G.; Gammon, D.; Katzer, D. S.; Park, D.; Sham, L. J. \textit{Science} \textbf{2000}, \textit{289} 1906.

\bibitem{crooker2} Crooker, S. A.; Barrick, T.; Hollingsworth, J. A.; Klimov, V. I. \textit{Appl. Phys. Lett.} \textbf{2003}, \textit{82} 2793.

\bibitem{gupta} Gupta, J. A.; Awschalom, D. D.; Peng, X.; Alivisatos, A. P. \textit{Phys. Rev. B} \textbf{1999}, \textit{59} R10421.

\bibitem{efros1} Efros, Al. L.; Rosen, M.; Kuno, M.; Nirmal, M.; Norris, D. J.; Bawendi, M. G. \textit{Phys. Rev. B} \textbf{1996}, \textit{54} 4843.

\bibitem{nirmal1} Nirmal, M.; Murray, C. B.; Bawendi, M. G. \textit{Phys. Rev. B} \textbf{1994}, \textit{50} 2293.

\bibitem{nirmal2} Nirmal, M.; Norris, D. J.; Kuno, M.; Bawendi, M. G. \textit{Phys. Rev. Lett.} \textbf{1995}, \textit{75} 3728.

\bibitem{chamarro} Chamarro, M.; Gourdon, C.; Lavallard, P. \textit{J. Lumin.} \textbf{1996}, \textit{70} 222.

\bibitem{murray} Murray, C. B.; Norris, D. J.; Bawendi, M. G. \textit{J. Am. Chem. Soc.} \textbf{1993}, \textit{115} 8706.

\bibitem{crooker3} Crooker, S. A.; Rickel, D. G.; Lyo, S. K.; Samarth, N.; Awschalom, D. D. \textit{Phys. Rev. B} \textbf{1999}, \textit{60} R2173.

\bibitem{efros2} Efros, Al. L., in \textit{Semiconductors and Metal Nanocrystals}, edited by V. I. Klimov (Marcel Dekker Inc., 2004).

\bibitem{johnstonhalperin} Johnston-Halperin, E.; Awschalom, D. D.; Crooker, S. A.; Efros, Al. L.; Rosen, M.; Peng, X.; Alivisatos, A. P. \textit{Phys. Rev. B} \textbf{2001}, \textit{63} 205309.

\bibitem{blackwood} Blackwood, E.; Snelling, M. J.; Harley, R. T.; Andrews, S. R.; Foxon, C. T. B. \textit{Phys. Rev. B} \textbf{1994}, \textit{50} 14246.

\bibitem{klimov2} Klimov, V. I.; McBranch, D. W.; Leatherdale, C. A.; Bawendi, M. G. \textit{Phys. Rev. B} \textbf{1999}, \textit{60} 13740.

\bibitem{achermann2} Achermann, M.; Petruska, M.; Crooker, S. A.; Klimov, V. I. \textit{J. Phys. Chem. B} \textbf{2003}, \textit{107} 13782.

\bibitem{kagan} Kagan, C. R.; Murray, C. B.; Nirmal, M.; Bawendi, M. G. \textit{Phys. Rev. Lett.} \textbf{1996}, \textit{76} 1517.

\bibitem{govorov} Govorov, A. O. \textit{Phys. Rev. B} \textbf{2003}, \textit{68} 075315.

\bibitem{ouyang} Ouyang, M.; Awschalom, D. D. \textit{Science} \textbf{2003}, \textit{301} 1074.


\end{references}
\end{document}